\documentclass[a4paper,11pt]{article}
\pdfoutput=1 

\usepackage{jheppub} 

\usepackage[T1]{fontenc} 

\usepackage{braket}

\title{\boldmath Entanglement Entropy of Disjoint Regions in Excited States
: An Operator Method}

\author{Noburo Shiba}

\affiliation{Yukawa Institute for Theoretical Physics (YITP),\\
Kyoto University, Kyoto 606-8502, Japan}

\abstract{
We develop the computational method of  
entanglement entropy based on the idea that 
$\mathrm{Tr} \rho_{\Omega}^n$  
is written  
as the expectation value of the local operator, 
where $\rho_{\Omega}$ is a density matrix of the subsystem $\Omega$. 
We apply it to consider the mutual R\'{e}nyi information 
$I^{(n)}(A,B)=S^{(n)}_A+S^{(n)}_B-S^{(n)}_{A\cup B}$ 
of disjoint compact spatial regions $A$ and $B$ in the locally excited states defined by acting 
the local operators at $A$ and $B$ on the vacuum 
of a $(d+1)$-dimensional field theory, 
in the limit when the separation $r$ between $A$ and $B$ is much greater than their sizes $R_{A,B}$.  
For the general QFT which has a mass gap, 
we compute $I^{(n)}(A,B)$ explicitly and find that this result is interpreted in terms of an entangled state 
in quantum mechanics. 
For a free massless scalar field, 
we show that for some classes of excited states, 
$I^{(n)}(A,B)-I^{(n)}(A,B)|_{r \rightarrow \infty} =C^{(n)}_{AB}/r^{\alpha (d-1)}$ 
where $\alpha=1~ \text{or}~ 2$ which is determined by the property of the local operators 
under the transformation $\phi \rightarrow  -\phi$ and $\alpha=2$ for the vacuum state. 
We give a method to compute $C^{(2)}_{AB}$ systematically. 

 }
 
\begin{document} 

\begin{flushright}
YITP-14-62
\end{flushright}

\maketitle
\flushbottom

\section{Introduction}
\label{intro}

The entanglement entropy in the quantum field theory (QFT) plays 
 important roles in many fields of physics such as the string theory, 
condensed matter physics, and the physics of the black hole. 
The entanglement entropy is a useful quantity which characterize 
quantum properties of  given states. 
For example, the entanglement entropy of ground states 
follows the area law \cite{Bombelli:1986rw, Sr, Ereview, La} if we consider a local quantum field theory with a UV fixed point,
while non-local field theories \cite{ShTa, Ka} or QFTs with fermi surfaces \cite{FS1} at UV cut off scale can violate the
area law.

For a given density matrix $\rho$ of the total system, 
the entanglement entropy of the subsystem $\Omega$ is defined as 
\begin{equation}
S_{\Omega} =-\mathrm{Tr} \rho_{\Omega} \ln \rho_{\Omega}, 
\end{equation}
where $\rho_{\Omega} =\mathrm{Tr}_{\Omega^{c}}\rho$ 
is the reduced density matrix of the subsystem $\Omega$ 
and $\Omega^c$ is the complement of $\Omega$. 
The R\'{e}nyi entropy $S_{\Omega}^{(n)}$ is defined as 
\begin{equation}
S_{\Omega}^{(n)} = \dfrac{1}{1-n} \ln \mathrm{Tr} \rho_{\Omega}^{n} .
\end{equation}
The limit $n\rightarrow 1$ coincides with the entanglement entropy 
$\lim_{n=1} S_{\Omega}^{(n)}=S_{\Omega}$.

In this paper we develop the computational method of R\'{e}nyi 
entanglement entropy based on the idea that 
$\mathrm{Tr} \rho_{\Omega}^n$ 
is written  
as the expectation value of the local operator at $\Omega$. 
We apply this method to consider the mutual R\'{e}nyi information 
$I^{(n)}(A,B)=S^{(n)}_A+S^{(n)}_B-S^{(n)}_{A\cup B}$ 
of disjoint compact spatial regions $A$ and $B$ in the locally excited states defined by acting 
the local operators at $A$ and $B$ on the vacuum 
of a $(d+1)$-dimensional field theory, 
in the limit when the separation $r$ between $A$ and $B$ is much greater than their sizes $R_{A,B}$. 

Our method is based on the idea that 
$\mathrm{Tr} \rho_{\Omega}^n$ 
is written  
as the expectation value of the local operator at $\Omega$. 
This idea was originally used to compute $I^{(n)}(A,B)$ in the vacuum state 
by Cardy \cite{Ca1}, Calabrese et al. \cite{Ca2}  and Headrick \cite{He}. 
We generalize this idea to an arbitrary density matrix $\rho$ and construct explicitly 
the local operator. 
The density matrix of the total system $\rho$ can be a mixed state and an excited state.  
We consider the general scalar field and do not specify its Hamiltonian. 
(Our method is applicable to QFT with interaction. )
We summarize our method. 
We consider $n$ copies of the scalar fields and the $j$-th copy of the scalar field is denoted by 
$\{ \phi^{(j)} \}$. 
Thus the total Hilbert space, $H^{(n)}$, is the tensor product of the $n$ copies of the Hilbert space, 
$H^{(n)}= H \otimes H \dots \otimes H$ where $H$ is the Hilbert space of one scalar field. 
We define the density matrix $\rho^{(n)}$ in  $H^{(n)}$ as 
\begin{equation}
\rho^{(n)} \equiv \rho \otimes \rho \otimes \dots \otimes \rho
\end{equation}
where $\rho$ is an arbitrary density matrix in $H$. 
We can express $\mathrm{Tr} \rho_{\Omega}^n$ as 
\begin{equation}
\mathrm{Tr} \rho_{\Omega}^n=\mathrm{Tr} (\rho^{(n)} E_{\Omega}),  \label{expectation intro}
\end{equation}
where 
\begin{equation}
\begin{split}
&E_{\Omega} =\int \prod_{j=1}^{n} \prod_{x \in \Omega} DJ^{(j)}(x) D K^{(j)}(x)  
\exp [i \int d^d x \sum_{l=1}^{n} J^{(l+1)}(x) \phi^{(l)}(x) ] \\
&\times  \exp [i\int d^d x \sum_{l=1}^{n} K^{(l)}(x) \pi^{(l)}(x) ]
 \times \exp [-i\int d^d x \sum_{l=1}^{n} J^{(l)} \phi^{(l)} ] ,
\end{split}  \label{opjkf intro}
\end{equation}
where $\pi(x)$ is a conjugate momenta of $\phi(x)$, $[\phi(x), \pi(y)]=i \delta^d (x-y)$, 
and $J^{(j)}(x)$ and $K^{(j)}(x)$ exist only in $\Omega$ and $J^{(n+1)}=J^{(1)}$ and 
we normalize the measure of the functional integral as 
$\int DJ^{(j) } \exp [i\int d^d x J^{(j)} (x) f(x) ] =\prod_{x \in \Omega} \delta (f(x)) $ 
where $f(x)$ is an arbitrary function. 
Notice that $\phi$ and $\pi$ in (\ref{opjkf intro}) are operators and the ordering is important. 
We call this operator $E_{\Omega}$ as  \textit{the glueing operator}. 
When $\rho$ is a pure state, $\rho=\ket{\Psi}\bra{\Psi}$, 
the equation (\ref{expectation intro}) becomes 
\begin{equation}
\mathrm{Tr} \rho_{\Omega}^n = \bra{\Psi^{(n)}} E_{\Omega} \ket{\Psi^{(n)}}  \label{formula intro}
\end{equation}
where 
\begin{equation}
\ket{\Psi^{(n)}} =\ket{\Psi}\ket{\Psi}\dots \ket{\Psi} .
\end{equation}
For a free scalar field, we can rewrite $E_{\Omega}$ in (\ref{opjkf intro}) 
using the normal ordering. 
In the case $n=2$, we obtain a simple expression of  $E_{\Omega}$ 
and reproduce the result that $I^{(2)}(A,B)$ in the vacuum state is proportional to the product of 
the electrostatic capacitance of each regions obtained by Cardy \cite{Ca1}.
Furthermore the simple expression of  $E_{\Omega}$ is useful for numerical calculation. 

The advantages of this method are that we can use ordinary technique in QFT such as 
OPE and the cluster decomposition property and 
that we can use the general properties and the explicit expression of the glueing operator to 
compute systematically the  R\'{e}nyi entropy for an arbitrary state.

We apply this method to the mutual R\'{e}nyi information 
$I^{(n)}(A,B)$ in the locally excited states. 
We consider the following locally excited state, 
\begin{equation}
\ket{\Psi} =N (O_{iA} O_{jB} +O_{i'A} O_{j'B} ) \ket{0}, 
\end{equation}
where $N$ is a real normalization constant and
 $O_{iA}$ and $O_{i'A}$ ($O_{jB}$ and $O_{j'B}$) are operators on $A$ ($B$) 
and $i$ and $i'$ ($j$ and $j'$) label a kind of operators. 
For 
the general QFT which has a mass gap, 
we compute $I^{(n)}(A,B)$ explicitly and find that this result is interpreted in terms of an entangled state 
in a quantum mechanical system which has finite degrees of freedom. 
For a free massless scalar field, 
we show that for some classes of excited states, 
$I^{(n)}(A,B)-I^{(n)}(A,B)|_{r \rightarrow \infty} =C^{(n)}_{AB}/r^{\alpha (d-1)}$ 
where $\alpha=1~ \text{or}~ 2$ which is determined by the property of the local operators 
under the transformation $\phi \rightarrow  -\phi$ and $\alpha=2$ for the vacuum state. 

The mutual information for a free scalar field in higher dimensions has been studied
in only a few papers \cite{CaHu1,CaHu2, Sh1, Sh2, Ca1, Sch, Her}. 
In these papers, the authors considered only the mutual information for the vacuum state.  
The relation between the mutual information and 
the physics of the black hole was considered in \cite{Sh1}.
Recently, it is proposed \cite{Fa} that the mutual information is obtained by 
the quantum correction to the holographic entanglement entropy formula \cite{RT}. 
It would be interesting to use our results to check this proposition. 

The entanglement entropy for an excited state 
defined by acting the local operator on the vacuum 
was considered in \cite{NNT, No, HNTW, CNT}. 
In \cite{NNT}, the subsystem is a half of the total space  
and the local operator exists at the complement of the subsystem 
and the time evolution of the entanglement entropy was considered. 
It was found that the entanglement entropy at late time 
is interpreted in terms of an entangled state 
in quantum mechanics \cite{NNT}. 
This result is analogous to our result 
in the general QFT which has a mass gap.  


The present paper is organized as follows. 
In section 2.1 we develop the computational method of the entanglement entropy. 
We derive the basic formula (\ref{expectation intro}) and  construct explicitly 
the glueing operator $E_{\Omega}$. 
In section 2.2 we investigate the general properties of $E_{\Omega}$. 
In section 3 we consider the mutual R\'{e}nyi information 
$I^{(n)}(A,B)$ in the locally excited states 
in the general QFT which has a mass gap. 
We compute $I^{(n)}(A,B)$ explicitly and find that this result is interpreted in terms of an entangled state 
in quantum mechanics. 
In section 4 we consider free scalar fields. 
In section 4.1 we rewrite $E_{\Omega}$ in (\ref{opjkf intro}) 
using the normal ordering. 
In the case $n=2$, we obtain a simple expression of  $E_{\Omega}$ and 
reproduce the result that $I^{(2)}(A,B)$ in the vacuum state is proportional to the product of 
the electrostatic capacitance of each regions obtained by Cardy. 
In section 4.2 we consider a free massless scalar field. 
We show that for some classes of excited states 
$I^{(n)}(A,B)-I^{(n)}(A,B)|_{r \rightarrow \infty} =C^{(n)}_{AB}/r^{\alpha (d-1)}$ 
where $\alpha=1~ \text{or}~ 2$ which is determined by the property of the local operators 
under the transformation $\phi \rightarrow  -\phi$ and $\alpha=2$ for the vacuum state. 
In section 5 we summarize our conclusion. 

\section{Operator formalism}
\label{formalism}

\subsection{Operator representation of $\mathrm{Tr} \rho_{\Omega}^n$ }

We represent the trace of the $n$th power of the reduced density matrix as the expectation value of the local
operator. 
As a model amenable to unambiguous calculation we deal with the scalar field
as a collection of coupled oscillators on a lattice of space points,
labeled by capital Latin indices, the displacement at each point giving the value of the scalar field there.
The local Hermitian variables $\hat{q}_A$ and $\hat{p}_B$ 
(coordinates and the conjugate momentum) obey the canonical commutation relations
\begin{equation}
[\hat{q}_A,\hat{p}_B]=i \delta_{AB}, ~~~ [\hat{q}_A,\hat{q}_B]=[\hat{p}_A,\hat{p}_B]=0 .
\end{equation}
The density matrix $\rho$ of the total system in coordinate representation is 
\begin{equation}
\rho ( q_A   ;  q'_B  ) = \bra{ \{ q_A \}  } \rho \ket{ \{q'_B \} }
\end{equation}
where $\{q_A \}$ denotes the collection of all $q_A$'s. 
We consider the arbitrary density matrix $\rho$.

Now consider a subsystem (or subregion) $\Omega$ in the space. 
The oscillators in this region will be specified by  lowercase Latin letters,
and those in its complement $\Omega ^c$ will be specified by Greek letters.
We can obtain a reduced density matrix
$\rho_{\Omega}$ for $\Omega$ by integrating out over $q^{\alpha }\in \mathbb{R}$ for each of the oscillators in $\Omega^c$, and then we have
\begin{equation}
\rho_{\Omega} (  q_a   ;  q'_b  ) = \int \prod_{\alpha} {dq_{\alpha}}  \rho ( q_a, q_{\alpha} ; q'_b ,q_{\alpha}  ) 
\end{equation}
We obtain the trace of the $n$th power of the reduced density matrix $\rho_{\Omega}$ as 
\begin{equation}
\begin{split}
&\mathrm{Tr} \rho_{\Omega}^n  = \int \prod_{j=1}^{n} \prod_{a} {dq_{a}^{(j)}}  
\rho_{\Omega} ( q_{a}^{(1)} ; q_{a}^{(2)}  ) 
\rho_{\Omega} ( q_{a}^{(2)} ; q_{a}^{(3)}  ) \dots 
\rho_{\Omega} ( q_{a}^{(n)} ; q_{a}^{(1)}  )  \\
&=\int \prod_{j=1}^{n} \prod_{a} {dq_{a}^{(j)}}  
 \prod_{\alpha} {dq_{\alpha}^{(j)}} 
\rho ( q_{a}^{(1)} , q_{\alpha}^{(1)}  ; q_{a}^{(2)} , q_{\alpha}^{(1)}  ) 
\rho ( q_{a}^{(2)} , q_{\alpha}^{(2)}  ; q_{a}^{(3)} , q_{\alpha}^{(2)}  ) \dots 
\rho ( q_{a}^{(n)} , q_{\alpha}^{(n)}  ; q_{a}^{(1)} , q_{\alpha}^{(n)}  )    .    
\end{split}
\end{equation}
We consider $n$ copies of the oscillators and the $j$-th copy of the oscillators is denoted by 
$\{ q_A^{(j)} \}$. 
Thus the total Hilbert space, $H^{(n)}$, is the tensor product of the $n$ copies of the Hilbert space, 
$H^{(n)}= H \otimes H \dots \otimes H$.
We define the following density matrix, 
\begin{equation}
\rho^{(n)} ( q_A^{(1)}, \dots ,  q_A^{(n)} ; q_B^{(1)'}, \dots , q_B^{(n)'}  )  \equiv 
\rho(q_A^{(1)}; q_B^{(1)'}) \rho(q_A^{(2)}; q_B^{(2)'}) \dots \rho(q_A^{(n)}; q_B^{(n)'}),
\end{equation}
i.e. 
\begin{equation}
\rho^{(n)} \equiv \rho \otimes \rho \otimes \dots \otimes \rho.
\end{equation}
Then we can rewrite $\mathrm{Tr} \rho_{\Omega}^n$ as 
\begin{equation}
\begin{split}
&\mathrm{Tr} \rho_{\Omega}^n  =\int \prod_{j=1}^{n} \prod_{a} {dq_{a}^{(j)}}  
 \prod_{\alpha} {dq_{\alpha}^{(j)}} 
 \prod_{b} {dq_{b}^{(j)'}}  
 \prod_{\beta} {dq_{\beta}^{(j)'}}  \\
&\times \rho ( q_{a}^{(1)} , q_{\alpha}^{(1)}  ; q_{b}^{(1)'} , q_{\beta}^{(1)'}  ) 
\rho ( q_{a}^{(2)} , q_{\alpha}^{(2)}  ; q_{b}^{(2)'} , q_{\beta}^{(2)'}  ) \dots 
\rho ( q_{a}^{(n)} , q_{\alpha}^{(n)}  ; q_{b}^{(n)'} , q_{\beta}^{(n)'}  )  \\
&\times \delta (q_{\alpha}^{(1)} -q_{\beta}^{(1)'} ) \delta (q_{\alpha}^{(2)} -q_{\beta}^{(2)'} ) \dots 
\delta (q_{\alpha}^{(n)} -q_{\beta}^{(n)'} )  \\ 
&\times \delta (q_{b}^{(1)'} -q_{a}^{(2)} ) \delta (q_{b}^{(2)'} -q_{a}^{(3)} ) \dots 
\delta (q_{b}^{(n)'} -q_{a}^{(1)} )     \\
&=\int  \prod_{j=1}^{n} \prod_{A} {dq_{A}^{(j)}}  
 \prod_{B} {dq_{B}^{(j)'}}  
 \rho^{(n)} ( q_A^{(1)}, \dots ,  q_A^{(n)} ; q_B^{(1)'}, \dots , q_B^{(n)'}  )   \\
&\times E_{\Omega}(  q_B^{(1)'}, \dots ,  q_B^{(n)'} ; q_A^{(1)}, \dots , q_A^{(n)} )  \\
&=\mathrm{Tr} (\rho^{(n)} E_{\Omega}) ,    
\end{split} \label{expectation}
\end{equation}
where 
\begin{equation}
\begin{split}
& E_{\Omega}(  q_B^{(1)'}, \dots ,  q_B^{(n)'} ; q_A^{(1)}, \dots , q_A^{(n)} )  
\equiv \bra{ \{ q_B^{(1)'} \} , \dots ,  \{ q_B^{(n)'} \} } E_{\Omega} 
\ket{ \{ q_A^{(1)} \} , \dots ,\{ q_A^{(n)}\}  } \\
&\equiv \prod_{a} \prod_{\alpha} \delta (q_{\alpha}^{(1)} -q_{\beta}^{(1)'} ) \delta (q_{\alpha}^{(2)} -q_{\beta}^{(2)'} ) \dots 
\delta (q_{\alpha}^{(n)} -q_{\beta}^{(n)'} )  
\times \delta (q_{b}^{(1)'} -q_{a}^{(2)} ) \delta (q_{b}^{(2)'} -q_{a}^{(3)} ) \dots 
\delta (q_{b}^{(n)'} -q_{a}^{(1)} )   .  \\
\end{split}  \label{op matrix elements}
\end{equation}
We call $E_{\Omega}$ as \textit{the glueing operator}. 
When $\rho$ is a pure state, $\rho=\ket{\Psi}\bra{\Psi}$, 
the equation (\ref{expectation}) becomes 
\begin{equation}
\mathrm{Tr} \rho_{\Omega}^n = \bra{\Psi^{(n)}} E_{\Omega} \ket{\Psi^{(n)}}  \label{formula}
\end{equation}
where 
\begin{equation}
\ket{\Psi^{(n)}} =\ket{\Psi}\ket{\Psi}\dots \ket{\Psi} .
\end{equation}

We represent $E_{\Omega}$ as a function of $\hat{q}$ and $\hat{p}$. 
We can rewrite $E_{\Omega}$ as 
\begin{equation}
\begin{split}
&E_{\Omega} =\int \prod_{j=1}^{n} \prod_{a} {dq_{a}^{(j)}}   
 \prod_{b} {dq_{b}^{(j)'}}   
 \ket{  \{ q_b^{(1)'} \}, \dots , \{ q_b^{(n)'}\} } \bra{ \{ q_a^{(1)} \}, \dots ,\{ q_a^{(n)} \} }  \\
&\times \delta (q_{b}^{(1)'} -q_{a}^{(2)} ) \delta (q_{b}^{(2)'} -q_{a}^{(3)} ) \dots 
\delta (q_{b}^{(n)'} -q_{a}^{(1)} )  \\
&=\int \prod_{j=1}^{n} \prod_{a} \dfrac{dJ_{a}^{(j)}}{2\pi}   
\exp [i(J_a^{(2)} \hat{q}_a^{(1)} + J_a^{(3)} \hat{q}_a^{(2)} +\dots +J_a^{(n)} \hat{q}_a^{(n-1)} 
+J_a^{(1)} \hat{q}_a^{(n)} )] \\
&\times \int \prod_{j=1}^{n} \prod_{a} {dq_{a}^{(j)}}   
 \prod_{b} {dq_{b}^{(j)'}}   
 \ket{ \{ q_b^{(1)'} \} , \dots , \{ q_b^{(n)'} \} } \bra{\{ q_a^{(1)} \}, \dots , \{ q_a^{(n)} \}  } \\
&\times \exp [-i(J_a^{(1)} \hat{q}_a^{(1)} + J_a^{(2)} \hat{q}_a^{(2)} +\dots +J_a^{(n)} \hat{q}_a^{(n)} )] ,
\end{split}  \label{opmiddle}
\end{equation}
where we have written the delta functions as the Fourier integrals. 
Note that $\hat{q}^{(l)}_a$ is a operator and not a integral variable. 
The middle term  in (\ref{opmiddle}) is the tensor product of the following operator,  
\begin{equation}
\int dq' \int dq \ket{q'}  \bra{q}= \int dK \exp [iK \hat{p}] \label{opk1},
\end{equation}
where we have omitted the subscripts for simplicity. 
We can check easily that (\ref{opk1}) is correct just by taking the matrix elements of its both sides. 
Thus we can rewrite the middle term in (\ref{opmiddle}) as
\begin{equation}
\begin{split}
&\int \prod_{j=1}^{n} \prod_{a} {dq_{a}^{(j)}}   
 \prod_{b} {dq_{b}^{(j)'}}   
 \ket{ \{ q_b^{(1)'} \}, \dots , \{ q_b^{(n)'} \} } \bra{\{ q_a^{(1)} \}, \dots , \{ q_a^{(n)} \} } \\
&=\int \prod_{j=1}^{n} \prod_{a} d K_{a}^{(j)}   
\exp [i(K_a^{(1)} \hat{p}_a^{(1)} + K_a^{(2)} \hat{p}_a^{(2)} +\dots +K_a^{(n)} \hat{p}_a^{(n)} )] .
\end{split} \label{opk}
\end{equation}
We substitute  (\ref{opk}) into (\ref{opmiddle}) and obtain 
\begin{equation}
\begin{split}
&E_{\Omega} =\int \prod_{j=1}^{n} \prod_{a \in \Omega} \dfrac{dJ_{a}^{(j)}}{2\pi} d K_{a}^{(j)}  
\exp [i(J_a^{(2)} \hat{q}_a^{(1)} + J_a^{(3)} \hat{q}_a^{(2)} +\dots +J_a^{(n)} \hat{q}_a^{(n-1)} 
+J_a^{(1)} \hat{q}_a^{(n)} )] \\
&\times  \exp [i(K_a^{(1)} \hat{p}_a^{(1)} + K_a^{(2)} \hat{p}_a^{(2)} +\dots +K_a^{(n)} \hat{p}_a^{(n)} )]
\\
&\times \exp [-i(J_a^{(1)} \hat{q}_a^{(1)} + J_a^{(2)} \hat{q}_a^{(2)} +\dots +J_a^{(n)} \hat{q}_a^{(n)} )] .
\end{split}  \label{opjk}
\end{equation}
From (\ref{opjk}), 
for the $(d+1)$ dimensional scalar field theory, 
we obtain 
\begin{equation}
\begin{split}
&E_{\Omega} =\int \prod_{j=1}^{n} \prod_{x \in \Omega} DJ^{(j)}(x) D K^{(j)}(x)  
\exp [i \int d^d x \sum_{l=1}^{n} J^{(l+1)}(x) \phi^{(l)}(x) ] \\
&\times  \exp [i\int d^d x \sum_{l=1}^{n} K^{(l)}(x) \pi^{(l)}(x) ]
 \times \exp [-i\int d^d x \sum_{l=1}^{n} J^{(l)} \phi^{(l)} ] .
\end{split}  \label{opjkf}
\end{equation}
where $\pi(x)$ is a conjugate momenta of $\phi(x)$, $[\phi(x), \pi(y)]=i \delta^d (x-y)$, and $J^{(j)}(x)$ and $K^{(j)}(x)$ exist only in $\Omega$ and $J^{(n+1)}=J^{(1)}$ and we normalize the measure of the functional integral as 
$\int DJ^{(j) } \exp [i\int d^d x J^{(j)} (x) f(x) ] =\prod_{x \in \Omega} \delta (f(x)) $ 
where $f(x)$ is an arbitrary function.

\subsection{General properties of the glueing operator $E_{\Omega}$}

We investigate some general properties of $E_{\Omega}$. 

(1) Symmetry: 
From (\ref{opjkf}), $E_{\Omega}$ is invariant under the sign changing transformation 
$\phi, \pi \rightarrow - \phi, -\pi$, i.e. 
\begin{equation}
E_{\Omega} (\phi^{(1)}, \dots , \phi^{(n)}, \pi^{(1)}, \dots , \pi^{(n)} ) =
E_{\Omega} (-\phi^{(1)}, \dots , -\phi^{(n)}, -\pi^{(1)}, \dots , -\pi^{(n)} )  . \label{property1}
\end{equation}

(2) Locality: 
When $\Omega=A\cup B$ and $A \cap B=\emptyset$, 
\begin{equation}
E_{A\cup B} = E_{A} E_{B}   . \label{property2}
\end{equation}

(3) 
For $n$ arbitrary operators $F_{j}$ $(j=1,2, \dots ,n)$ on $H$, 
\begin{equation}
\mathrm{Tr} (F_1 \otimes F_2 \otimes \dots \otimes F_n \cdot  E_{\Omega} ) 
= \mathrm{Tr} ( F_{1\Omega} F_{2\Omega} \dots F_{n\Omega}   ) , \label{property3}
\end{equation}
where $F_{j\Omega} \equiv \mathrm{Tr}_{\Omega^c} F_{j} $. 
This property is the generalization of (\ref{expectation}).  
We can prove easily (\ref{property3}) by using the matrix elements of $E_{\Omega}$ 
in (\ref{op matrix elements}).

(4) The cyclic property: 
From the cyclic property of the trace in the right hand side in (\ref{property3}), we obtain 
\begin{equation}
\mathrm{Tr} (F_1 \otimes F_2 \otimes \dots \otimes F_n \cdot  E_{\Omega} ) 
= \mathrm{Tr} (F_2 \otimes F_3 \otimes \dots \otimes F_n \otimes F_1 \cdot  E_{\Omega} )   \label{property4}.
\end{equation}

(5) The relation between $E_{\Omega}$ and $E_{\Omega^c}$ for pure states: 
We consider two pure states $\ket{\phi_1} \ket{\phi_2} \dots \ket{\phi_n}$ and 
$\ket{\psi_1} \ket{\psi_2} \dots \ket{\psi_n}$ in $H^{(n)}$ where $\ket{\phi_j}$ and $\ket{\psi_j}$ $(j=1,2,\dots ,n)$ are arbitrary pure states. 
We can prove the following equation:
\begin{equation}
\bra{\psi_1} \bra{\psi_2} \dots \bra{\psi_n} E_{\Omega} 
\ket{\phi_1} \ket{\phi_2} \dots \ket{\phi_n} 
=[\bra{\phi_2} \bra{\phi_3} \dots \bra{\phi_n} \bra{\phi_1} E_{\Omega^c} 
\ket{\psi_1} \ket{\psi_2} \dots \ket{\psi_n} ]^*
  \label{property5}.
\end{equation}
We prove (\ref{property5}) as follows: 

From (\ref{op matrix elements}) we obtain
\begin{equation}
\begin{split}
&\bra{\psi_1} \bra{\psi_2} \dots \bra{\psi_n} E_{\Omega} 
\ket{\phi_1} \ket{\phi_2} \dots \ket{\phi_n}   
=\int \prod_{j=1}^{n} \prod_{a} {dq_{a}^{(j)}}  
 \prod_{\alpha} {dq_{\alpha}^{(j)}}   \\
&\times \psi_1 ( q_{a}^{(1)} , q_{\alpha}^{(1)}   )^* 
\psi_2 ( q_{a}^{(2)} , q_{\alpha}^{(2)}  )^* \dots 
\psi_n ( q_{a}^{(n)} , q_{\alpha}^{(n)}  )^*  
  \phi_1 ( q_{a}^{(n)} , q_{\alpha}^{(1)}   ) 
\phi_2 ( q_{a}^{(1)} , q_{\alpha}^{(2)}  ) \dots 
\phi_n ( q_{a}^{(n-1)} , q_{\alpha}^{(n)}  )    \\
&=[\int \prod_{j=1}^{n} \prod_{a} {dq_{a}^{(j)}}  
 \prod_{\alpha} {dq_{\alpha}^{(j)}}   
 \phi_2 ( q_{a}^{(1)} , q_{\alpha}^{(2)}   )^* 
\phi_3 ( q_{a}^{(2)} , q_{\alpha}^{(3)}  )^* \dots 
\phi_n ( q_{a}^{(n-1)} , q_{\alpha}^{(n)}  )^*  \phi_1 ( q_{a}^{(n)} , q_{\alpha}^{(1)}   )^* \\
&\times \psi_1 ( q_{a}^{(1)} , q_{\alpha}^{(1)}   ) 
\psi_2 ( q_{a}^{(2)} , q_{\alpha}^{(2)}  ) \dots 
\psi_n ( q_{a}^{(n)} , q_{\alpha}^{(n)}  )]^*  \\
&=[\bra{\phi_2} \bra{\phi_3} \dots \bra{\phi_n} \bra{\phi_1} E_{\Omega^c} 
\ket{\psi_1} \ket{\psi_2} \dots \ket{\psi_n} ]^*.  ~~~~ \square
\end{split} \label{property5-2}
\end{equation}

By using (\ref{property5}), we can obtain the following basic property 
 (see e.g. \cite{Ni}) for a pure state 
$\rho=\ket{\Psi}\bra{\Psi}$;
\begin{equation}
\mathrm{Tr} \rho_{\Omega}^n = \bra{\Psi^{(n)}} E_{\Omega} \ket{\Psi^{(n)}}
=[ \bra{\Psi^{(n)}} E_{\Omega^c} \ket{\Psi^{(n)}} ]^* 
=\mathrm{Tr} \rho_{\Omega^c}^n .
\end{equation}
where we have used the fact that $\mathrm{Tr} \rho_{\Omega}^n$ is real. 
So  (\ref{property5}) is the generalization of 
$\mathrm{Tr} \rho_{\Omega}^n =\mathrm{Tr} \rho_{\Omega^c}^n $ for a pure state 
$\rho=\ket{\Psi}\bra{\Psi}$.

\section{Locally excited states in the general QFT which has a mass gap}
We consider the mutual R\'{e}nyi information  
$I^{(n)}(A,B) \equiv S_{A}^{(n)} +S_{B}^{(n)} -S_{A\cup B}^{(n)}$ of
disjoint compact spatial regions A and B in the locally excited states of 
the general QFT which has the mass gap $m$ in the limit when the separation $r$ between $A$ and
$B$ is much greater than their sizes $R_{A,B}$ and $1/m$ ($r \gg R_{A,B}, 1/m$).
We consider the following locally excited state, 
\begin{equation}
\ket{\Psi} =N (O_{iA} O_{jB} +O_{i'A} O_{j'B} ) \ket{0}, 
\end{equation}
where $N$ is a real normalization constant and
 $O_{iA}$ and $O_{i'A}$ ($O_{jB}$ and $O_{j'B}$) are operators on $A$ ($B$) 
and $i$ and $i'$ ($j$ and $j'$) label a kind of operators. 
We assume the distance between the positions of  $O_{A}$   ($O_{B}$) 
 and the boundary of $A$ ($B$) is much greater than $1/m$ ($ R_{A,B} \gg 1/m$) 
(we have omitted the labels $i$s for simplicity.) 
We impose following orthogonal conditions for simplicity, 
\begin{equation}
\bra{0} O_{iA}^{\dagger} O_{i'A} \ket{0}  =\bra{0} O_{jB}^{\dagger} O_{j'B} \ket{0}=0 . \label{massive orthogonal}
\end{equation}
This state is similar to the EPR state. 
We compute the mutual R\'{e}nyi information of this state.  
In this calculation, the general properties (2), (5) 
and the cluster decomposition property in the QFT play important roles.

From the normalization condition $\braket{\Psi|\Psi}=1$ we obtain 
\begin{equation}
N^{-2} \simeq \bra{0} O_{iA}^{\dagger} O_{iA} \ket{0}  \bra{0} O_{jB}^{\dagger} O_{jB} \ket{0} 
+ \bra{0} O_{i'A}^{\dagger} O_{i'A} \ket{0}  \bra{0} O_{j'B}^{\dagger} O_{j'B} \ket{0} , 
\label{normalization mass gap}
\end{equation}
where we have used the cluster decomposition property 
$\bra{0} O_{A}^{\dagger} O_{A} O_{B}^{\dagger} O_{B} \ket{0} \simeq
\bra{0} O_{A}^{\dagger} O_{A} \ket{0}  \bra{0} O_{B}^{\dagger} O_{B} \ket{0}$
and the orthogonal conditions  (\ref{massive orthogonal}).  
By using (\ref{formula}), we obtain $\mathrm{Tr} \rho_{\Omega}^n$ for $\rho=\ket{\Psi}\bra{\Psi}$ as 
\begin{equation}
\begin{split}
&\mathrm{Tr} \rho_{\Omega}^n = \bra{\Psi^{(n)}} E_{\Omega} \ket{\Psi^{(n)}}  \\
&=N^{2n} \bra{0^{(n)}}  (O_{iA}^\dagger O_{jB}^\dagger +O_{i'A}^\dagger O_{j'B}^\dagger ) ^{(1)} \dots 
 (O_{iA}^\dagger O_{jB}^\dagger +O_{i'A}^\dagger O_{j'B}^\dagger ) ^{(n)}  \\
& \times E_{\Omega} 
  (O_{iA} O_{jB} +O_{i'A} O_{j'B} ) ^{(1)} \dots 
 (O_{iA} O_{jB} +O_{i'A} O_{j'B} ) ^{(n)} \ket{0^{(n)}} , \label{omega mass gap}
\end{split}
\end{equation}
where $O^{(l)}$ are operators in the Hilbert space of the $l$-th copy 
and the $\Omega$ is $A, B$ or $A\cup B$.  

First, we consider $\mathrm{Tr} \rho_{A\cup B}^n$. 
From (\ref{property5}) and (\ref{omega mass gap}) we obtain 
\begin{equation}
\begin{split}
&\mathrm{Tr} \rho_{A \cup B}^n  
=N^{2n} [\bra{0^{(n)}}  (O_{iA}^\dagger O_{jB}^\dagger +O_{i'A}^\dagger O_{j'B}^\dagger ) ^{(1)} \dots 
 (O_{iA}^\dagger O_{jB}^\dagger +O_{i'A}^\dagger O_{j'B}^\dagger ) ^{(n)}  \\
& \times E_{(A \cup B)^c} 
  (O_{iA} O_{jB} +O_{i'A} O_{j'B} ) ^{(1)} \dots 
 (O_{iA} O_{jB} +O_{i'A} O_{j'B} ) ^{(n)} \ket{0^{(n)}}]^* \\
& \simeq \braket{\Psi | \Psi}   [\bra{0^{(n)}} E_{(A \cup B)^c}  \ket{0^{(n)}}]^* 
=\bra{0^{(n)}} E_{A \cup B}  \ket{0^{(n)}} 
=\bra{0^{(n)}} E_{A} E_{B}  \ket{0^{(n)}} \\
& \simeq  \bra{0^{(n)}} E_{A }  \ket{0^{(n)}} \bra{0^{(n)}} E_{B }  \ket{0^{(n)}} 
= \mathrm{Tr} \rho_{0A }^n  \mathrm{Tr} \rho_{0B }^n   , \label{AB mass gap}
\end{split}
\end{equation}
where we have used the cluster decomposition property and the conditions $r,R_A,R_B \gg 1/m$ 
and  $\rho_{0A(B) }$ is the reduced density matrix of the vacuum state.  

Next we consider $\mathrm{Tr} \rho_{A}^n$. 
We expand the product in (\ref{omega mass gap}). 
The terms in the expansion in (\ref{omega mass gap}) have the following form,  
\begin{equation}
\begin{split}
& \bra{0^{(n)}}  O_{i_1 A}^{(1)\dagger} \dots  O_{i_n A}^{(n)\dagger}  E_{A } 
O_{i_{n+1} A}^{(1)} \dots  O_{i_{2n} A}^{(n)} 
\cdot O_{j_1 B}^{(1)\dagger} \dots  O_{j_n B}^{(n)\dagger}   
O_{j_{n+1} B}^{(1)} \dots  O_{j_{2n} B}^{(n)}  \ket{0^{(n)}} \\
& \simeq  \bra{0^{(n)}}  O_{i_1 A}^{(1)\dagger} \dots  O_{i_n A}^{(n)\dagger}  E_{A } 
O_{i_{n+1} A}^{(1)} \dots  O_{i_{2n} A}^{(n)}  \ket{0^{(n)}} 
\bra{0^{(n)}} O_{j_1 B}^{(1)\dagger} \dots  O_{j_n B}^{(n)\dagger}   
O_{j_{n+1} B}^{(1)} \dots  O_{j_{2n} B}^{(n)}  \ket{0^{(n)}} \\
& = \bra{0^{(n)}}  O_{i_1 A}^{(1)\dagger} \dots  O_{i_n A}^{(n)\dagger}  E_{A } 
O_{i_{n+1} A}^{(1)} \dots  O_{i_{2n} A}^{(n)}  \ket{0^{(n)}} 
\prod_{l=1}^{n} \bra{0}  O_{j_l B}^{\dagger} O_{j_{l+n} B} \ket{0} \\
& = \bra{0^{(n)}}  O_{i_1 A}^{(1)\dagger} \dots  O_{i_n A}^{(n)\dagger}  E_{A } 
O_{i_{n+1} A}^{(1)} \dots  O_{i_{2n} A}^{(n)}  \ket{0^{(n)}} 
\prod_{l=1}^{n} \delta_{j_l j_{l+n}} \bra{0}  O_{j_l B}^{\dagger} O_{j_{l} B} \ket{0}
\end{split}  \label{term in expansion}
\end{equation}
where $(i_l, j_l)$ is $(i,j)$ or $(i',j')$ $(l=1,2,\dots , 2n)$ and  
we have used the cluster decomposition property and the condition $r \gg 1/m$ in the second line 
and used the orthogonal conditions in (\ref{massive orthogonal}) in the last line. 
From (\ref{term in expansion}), the number of the nonzero terms in the 
the expansion in (\ref{omega mass gap}) is $2^n$. 
From (\ref{omega mass gap}) and  (\ref{term in expansion}), we obtain 
\begin{equation}
\begin{split}
&\mathrm{Tr} \rho_{A}^n  
=N^{2n} [  \bra{0}  O_{j B}^{\dagger} O_{j B} \ket{0}^{n} \bra{0^{(n)}}  
O_{i A}^{(1)\dagger} \dots  O_{i A}^{(n)\dagger}  E_{A } 
O_{i A}^{(1)} \dots  O_{i A}^{(n)}   \ket{0^{(n)}} \\
& +  \bra{0}  O_{j B}^{\dagger} O_{j B} \ket{0}^{n-1} \bra{0}  O_{j' B}^{\dagger} O_{j' B} \ket{0} 
(\bra{0^{(n)}}  O_{i' A}^{(1)\dagger}  O_{i A}^{(2)\dagger} \dots  O_{i A}^{(n)\dagger}  E_{A } 
O_{i' A}^{(1)} O_{i A}^{(2)}  \dots  O_{i A}^{(n)}   \ket{0^{(n)}} + \dots )  \\
&+ \bra{0}  O_{j B}^{\dagger} O_{j B} \ket{0}^{n-2} \bra{0}  O_{j' B}^{\dagger} O_{j' B} \ket{0}^{2}
 (\bra{0^{(n)}}  O_{i' A}^{(1)\dagger}  O_{i' A}^{(2)\dagger} O_{i A}^{(3)\dagger} \dots  O_{i A}^{(n)\dagger} 
  E_{A } 
O_{i' A}^{(1)} O_{i' A}^{(2)} O_{i A}^{(3)} \dots  O_{i A}^{(n)}   \ket{0^{(n)}} + \dots )  \\
&+ \dots +
\bra{0}  O_{j' B}^{\dagger} O_{j' B} \ket{0}^{n} \bra{0^{(n)}}  
O_{i' A}^{(1)\dagger} \dots  O_{i' A}^{(n)\dagger}  E_{A }  
O_{i' A}^{(1)} \dots  O_{i' A}^{(n)}   \ket{0^{(n)}} ]  .
\end{split}  \label{A mass gap1}
\end{equation}
The number of the terms which are proportional to 
$\bra{0}  O_{j B}^{\dagger} O_{j B} \ket{0}^{n-l} \bra{0}  O_{j' B}^{\dagger} O_{j' B} \ket{0}^{l} $ 
in (\ref{A mass gap1}) 
is ${}_n \mathrm{C}_l$. 
By using (\ref{property5}), we obtain 
\begin{equation}
\begin{split}
& \bra{0^{(n)}}  O_{i_1 A}^{(1)\dagger} \dots  O_{i_n A}^{(n)\dagger}  E_{A } 
O_{i_{1} A}^{(1)} \dots  O_{i_{n} A}^{(n)}   \ket{0^{(n)}} \\
& = [\bra{0^{(n)}}  O_{i_2 A}^{(1)\dagger} \dots  O_{i_n A}^{(n-1)\dagger} 
O_{i_1 A}^{(n)\dagger}  E_{A^c } 
O_{i_{1} A}^{(1)} \dots  O_{i_{n} A}^{(n)}   \ket{0^{(n)}} ]^*  \\
& \simeq [\bra{0^{(n)}} E_{A^c}  \ket{0^{(n)}} 
\bra{0^{(n)}}  O_{i_2 A}^{(1)\dagger} \dots  O_{i_n A}^{(n-1)\dagger} 
O_{i_1 A}^{(n)\dagger}  
O_{i_{1} A}^{(1)} \dots  O_{i_{n} A}^{(n)}   \ket{0^{(n)}}   ]^* \\
&=\bra{0^{(n)}} E_{A}  \ket{0^{(n)}} 
\prod_{l=1}^n \bra{0} O_{i_{l}A}^{\dagger} O_{i_{l+1}A} \ket{0} \\
&= 
\begin{cases}
\mathrm{Tr} \rho_{0A}^n \bra{0}  O_{iA}^{\dagger} O_{iA} \ket{0}^{n}  &\text{for}~ i= i_1=i_2=\dots=i_n  \\
\mathrm{Tr} \rho_{0A}^n \bra{0}  O_{i'A}^{\dagger} O_{i'A} \ket{0}^{n}  &\text{for}~ i'= i_1=i_2=\dots=i_n  \\
0  &\text{otherwise}  \\
\end{cases}
\end{split}  \label{term in expansion2}
\end{equation}
where $i_{n+1}=i_1$ and 
we have used the cluster decomposition property and the condition $R_{A} \gg 1/m$ 
in the third line 
and used the orthogonal conditions in (\ref{massive orthogonal}) in the last line. 
From (\ref{term in expansion2}) the nonzero terms in (\ref{A mass gap1}) are only 
the first term and the last term and we obtain 
\begin{equation}
\begin{split}
&\mathrm{Tr} \rho_{A}^n  
=\mathrm{Tr} \rho_{0A}^n \cdot N^{2n} 
[\bra{0} O_{iA}^{\dagger} O_{iA} \ket{0}^{n} \bra{0}  O_{j B}^{\dagger} O_{j B}\ket{0}^{n} 
+\bra{0} O_{i'A}^{\dagger} O_{i'A} \ket{0}^{n} \bra{0}  O_{j' B}^{\dagger} O_{j' B} \ket{0}^{n} ]  .
\end{split}  \label{A mass gap2}
\end{equation}
By the same way, we obtain $\mathrm{Tr} \rho_{B}^n$ as 
\begin{equation}
\begin{split}
&\mathrm{Tr} \rho_{B}^n  
=\mathrm{Tr} \rho_{0B}^n \cdot N^{2n} 
[\bra{0} O_{iA}^{\dagger} O_{iA} \ket{0}^{n} \bra{0}  O_{j B}^{\dagger} O_{j B}\ket{0}^{n} 
+\bra{0} O_{i'A}^{\dagger} O_{i'A} \ket{0}^{n} \bra{0}  O_{j' B}^{\dagger} O_{j' B} \ket{0}^{n} ]  .
\end{split}  \label{B mass gap2}
\end{equation}
From (\ref{normalization mass gap}), (\ref{AB mass gap}), (\ref{A mass gap2}) and  (\ref{B mass gap2}), 
we obtain the mutual R\'{e}nyi information  
$I^{(n)}(A,B) =(n-1)^{-1} \ln \tfrac{\mathrm{Tr} \rho_{A \cup B}^n}{\mathrm{Tr} \rho_{A}^n \mathrm{Tr} \rho_{B}^n}$ as
\begin{equation}
I^{(n)}(A,B)=\dfrac{2}{n-1} \ln \dfrac{(x+y)^n}{x^n+y^n}
\label{mutual Renyi mass gap}
\end{equation}
where 
\begin{equation}
x\equiv \bra{0} O_{iA}^{\dagger} O_{iA} \ket{0} \bra{0}  O_{j B}^{\dagger} O_{j B}\ket{0} , 
~~y\equiv\bra{0} O_{i'A}^{\dagger} O_{i'A} \ket{0} \bra{0}  O_{j' B}^{\dagger} O_{j' B}\ket{0}. 
\end{equation}
Taking the limit $n\rightarrow 1$ leads to the the mutual information  
\begin{equation}
I(A,B)=2 [\ln (x+y)-\dfrac{1}{x+y}(x\ln x+y\ln y)].
\label{mutual mass gap}
\end{equation}

We can reproduce these results from the quantum mechanics. 
Let us consider the following state, 
\begin{equation}
\ket{\Psi}_{qm} =N (\ket{i}_A \ket{j}_B +\ket{i'}_A \ket{j'}_B  ) , \label{state qm} 
\end{equation}
where $N$ is the real normalization constant, and 
 $\ket{i(i')}_A$ and  $\ket{j(j')}_B$ are the pure state of the subsystem $A$ and $B$ and 
\begin{equation}
\braket{i|i'}_A=\braket{j|j'}_B=0 . \label{orthogonal qm}
\end{equation}
From (\ref{state qm}), (\ref{orthogonal qm}) and the normalization condition 
$\braket{\Psi|\Psi}_{qm}=1$, we obtain   
\begin{equation}
\mathrm{Tr} \rho_{qmA}^n=\mathrm{Tr} \rho_{qmB}^n 
=N^{2n} [\braket{i|i}_A^n \braket{j|j}_B^n + \braket{i'|i'}_A^n \braket{j'|j'}_B^n ] 
\label{Aqm}
\end{equation}
where 
\begin{equation}
N^{-2}=\braket{i|i}_A \braket{j|j}_B + \braket{i'|i'}_A \braket{j'|j'}_B . \label{normalization qm}
\end{equation}
Because the total system is a pure state, $\mathrm{Tr} \rho_{qm A \cup B}^n=1$. 
From (\ref{Aqm}) and (\ref{normalization qm}) we obtain the mutual R\'{e}nyi information as 
\begin{equation}
I_{qm}^{(n)}(A,B)=\dfrac{2}{n-1} \ln \dfrac{(x_{qm}+y_{qm})^n}{x_{qm}^n+y_{qm}^n} ,
\label{mutual Renyi qm}
\end{equation}
where 
\begin{equation}
x_{qm}\equiv \braket{i|i}_A \braket{j|j}_B , 
~~y_{qm}\equiv \braket{i'|i'}_A \braket{j'|j'}_B   . 
\end{equation}
(\ref{mutual Renyi mass gap}) is the same as (\ref{mutual Renyi qm})  
when we replace the states as follows: 
\begin{equation}
O_{i(i')A}\ket{0} \rightarrow  \ket{i(i')}_A, ~~ O_{j(j')B}\ket{0} \rightarrow  \ket{j(j')}_B. 
\end{equation}
Interestingly, the mutual information in the QFT measures only the quantum entanglement  
in the limit $r \rightarrow \infty$ although 
the mutual information measures generally 
the total of the quantum entanglement and the classical one \cite{Ni}. 
So, in this limit, the mutual information is a good measure of quantum entanglement in this sense.


\section{Free scalar fields}
\subsection{Explicit calculation of the glueing operator $E_{\Omega}$}
We consider  $(d+1)$ dimensional free scalar field theory. 
For free scalar fields, it is useful to represent the glueing operator $E_{\Omega}$ 
in (\ref{opjkf}) as the normal ordered operator. 
We decompose $\phi$ and $\pi$ into the creation and annihilation parts, 
\begin{equation}
\phi (x) =\phi^{+}(x) +\phi^{-}(x), ~~~\pi (x) =\pi^{+}(x) +\pi^{-}(x),
\end{equation}
where 
\begin{equation}
\begin{split}
&\phi^{+} (x) = \int \dfrac{d^d p}{(2\pi)^d} \dfrac{1}{\sqrt{2E_p}} a_p e^{ipx}, 
~~\phi^{-} (x)=(\phi^{+} (x))^{\dagger} ,  \\
&\pi^{+} (x) = \int \dfrac{d^d p}{(2\pi)^d} (-i) \sqrt{\dfrac{E_p}{2}} a_p e^{ipx}, 
~~\pi^{-} (x)=(\pi^{+} (x))^{\dagger}  ,
\end{split}
\end{equation}
here  $E_p$ is the energy and  $[a_p,a_{p'}^{\dagger}] =(2\pi)^d \delta^{d} (p-p')$. 
The commutators of these operators are 
\begin{equation}
\begin{split}
&[\phi^{+} (x), \phi^{-} (y)]  = \bra{0} \phi (x) \phi (y) \ket{0} 
= \int \dfrac{d^d p}{(2\pi)^d} \dfrac{1}{2E_p}  e^{ip(x-y)} \equiv \dfrac{1}{2} W^{-1}(x-y),   \\
&[\pi^{+} (x), \pi^{-} (y)]  = \bra{0} \pi (x) \pi (y) \ket{0} 
= \int \dfrac{d^d p}{(2\pi)^d} \dfrac{E_p}{2}  e^{ip(x-y)} \equiv \dfrac{1}{2} W(x-y),          \\
&[\pi^{+} (x), \phi^{-} (y)] =[\pi^{-} (x), \phi^{+} (y)] = -\dfrac{i}{2} \delta^d (x-y)  ,
\end{split}  \label{commutators}
\end{equation}
where 
 we have defined the matrix $W$ which has continuous indices $x,y$ in (\ref{commutators}) 
and $W^{-1}$ is the inverse of $W$. 
By using (\ref{commutators}) and the Baker-Campbell-Hausdorff (BCH) formula 
$e^X e^Y =e^{[X,Y]} e^Y e^X, ~~ e^{X+Y} = e^{-\frac{1}{2}[X,Y] } e^X e^Y$,  
for $[[X, Y], X] = [[X, Y], Y]=0$, 
we obtain 
\begin{equation}
\begin{split}
&\exp [i\int d^d x J' \phi ] \exp [i\int d^d x K \pi ]  \exp [-i\int d^d x J \phi ]  \\
&=:\exp [i\int d^d x (K \pi + (J'-J)\phi)] :   \\
&\times \exp[\int d^d x d^d y (-\dfrac{1}{4} K(x) W(x-y)K(y) -\dfrac{1}{4} (J'-J)(x) W^{-1}(x-y)(J'-J)(y) )  \\
&-\int d^dx \dfrac{i}{2} K(x) (J'+J)(x)  ] ,  \\
\end{split}  \label{normal order}
\end{equation}
where $:O:$ means the normal ordered operator of $O$. 
From (\ref{normal order}) 
we can rewrite $E_{\Omega}$  in (\ref{opjkf}) as 
the normal ordered operator,  
\begin{equation}
\begin{split}
&E_{\Omega} =\int \prod_{j=1}^{n} \prod_{x \in \Omega} DJ^{(j)}(x) D K^{(j)}(x)  
:\exp [i \sum_{l=1}^{n} \int d^d x ( (J^{(l+1)} -J^{(l)} ) \phi^{(l)} +K^{(l)} \pi^{(l)} ) ]: \exp [- \tilde{S}] , \\
\end{split}  \label{normal opjkf}
\end{equation}
where $J^{(n+1)} =J^{(1)}$ and 
\begin{equation}
\begin{split}
& \tilde{S} \equiv  \sum_{l=1}^{n} [\int d^d x d^d y [\dfrac{1}{4} K^{(l)}(x) W(x-y) K^{(l)}(y) 
+\dfrac{1}{4} (J^{(l+1)}- J^{(l)} ) (x) W^{-1}(x-y) (J^{(l+1)}- J^{(l)} )(y)]  \\
&+\dfrac{i}{2} \int d^d x K^{(l)}(x) (J^{(l+1)}+J^{(l)})(x) ] . 
\end{split}  \label{tilde S}
\end{equation}
For the vacuum state $\rho_{0}=\ket{0}\bra{0}$, 
from (\ref{normal opjkf}) we obtain 
\begin{equation}
\mathrm{Tr} \rho_{0\Omega}^n = \bra{0^{(n)}} E_{\Omega} \ket{0^{(n)}}  
=\int \prod_{j=1}^{n} \prod_{x \in \Omega} DJ^{(j)}(x) D K^{(j)}(x)  \exp [- \tilde{S}] . \label{vacuum}
\end{equation}
We can show that (\ref{vacuum}) reproduces the same result as that of 
the real time approach \cite{Bombelli:1986rw, Sr, CW}
 which is based on the wave functional calculation (see Appendix A).  
By expanding the exponential in the normal ordered product in (\ref{normal opjkf}) 
and performing the Gauss integral 
of $J$ and $K$, we can rewrite the $E_{\Omega}$ as a series of operators.  
Note that the odd powers of the expansion of (\ref{normal opjkf}) vanish 
from the symmetric property (\ref{property1}).

\subsubsection{The case $n=2$}
In the case $n=2$, it is useful to define the following linear combinations, 
\begin{equation}
\phi_{\pm}=\dfrac{1}{\sqrt{2}} (\phi^{(1)} \pm \phi^{(2)}), 
~~\pi_{\pm}=\dfrac{1}{\sqrt{2}} (\pi^{(1)} \pm \pi^{(2)}), 
~~J_{\pm}=\dfrac{1}{\sqrt{2}} (J^{(1)} \pm J^{(2)}),
~~K_{\pm}=\dfrac{1}{\sqrt{2}} (K^{(1)} \pm K^{(2)}).  \label{linear combinations}
\end{equation}
From (\ref{linear combinations}) we rewrite (\ref{normal opjkf}) as 
\begin{equation}
\begin{split}
&E_{\Omega} =\int DJ_{+} DJ_{-} DK_{+} DK_{-}  
:\exp [i \int d^d x (-2J_{-} \phi_{-} +K_{+} \pi_{+} +K_{-} \pi_{-} ) ]:  \\
&\times  \exp [  \int d^d x d^d y (-\dfrac{1}{4} (K_{+}(x) W(x-y) K_{+}(y) + K_{-}(x) W(x-y) K_{-}(y) ) 
-J_{-} (x) W^{-1}(x-y) J_{-}(y)  ) \\
&-i\int d^d x K_{+}(x) J_{+}(x)  ] \\
&=\int DJ_{-} DK_{-}  
:\exp [i \int d^d x (-2J_{-} \phi_{-} +K_{-} \pi_{-} ) ]:  \\
&\times  \exp [  \int d^d x d^d y (-\dfrac{1}{4} K_{-}(x) W(x-y) K_{-}(y)  
-J_{-} (x) W^{-1}(x-y) J_{-}(y)  ) ]  ,         \\
\end{split}  \label{normal opjkf n=2}
\end{equation}
where we have performed $J_{+}$ and $K_{+}$ integrals. 
In order to represent the Gauss integrals of  $K_{-}$ and  $J_{-}$, 
we will use the following matrix notation, 
\begin{alignat}{2}
W(x-y) = \begin{pmatrix} 
             W(x_\Omega - y_\Omega) & W(x_\Omega - y_{\Omega^c})  \\
             W(x_{\Omega^c} - y_\Omega) & W(x_{\Omega^c} - y_{\Omega^c}) 
             \end{pmatrix} 
             \equiv   \begin{pmatrix} 
             A & B  \\
             B^T & C 
             \end{pmatrix}  &   ~~~~~~ \label{matw} \\
 W^{-1}(x-y) = \begin{pmatrix} 
             W^{-1}(x_\Omega - y_\Omega) & W^{-1}(x_\Omega - y_{\Omega^c})  \\
             W^{-1}(x_{\Omega^c} - y_\Omega) & W^{-1}(x_{\Omega^c} - y_{\Omega^c}) \end{pmatrix}
              \equiv   \begin{pmatrix} 
             D & E  \\
             E^T & F 
             \end{pmatrix}             
 \label{matwin}
\end{alignat}
where $x_{\Omega (\Omega^c)}$ and $y_{\Omega (\Omega^c)}$ are the coordinates 
in $\Omega (\Omega^c)$. 
Thus the propagators of $J_{-}$ and $K_{-}$ are  
\begin{equation}
<J_{-}(x) J_{-}(y)> \equiv \dfrac{\int D J_{-} J_{-}(x) J_{-}(y) 
e^{ -\int d^d x d^d y J_{-} (x) W^{-1}(x-y) J_{-}(y) }  }{\int D J_{-}  
e^{ -\int d^d x d^d y J_{-} (x) W^{-1}(x-y) J_{-}(y)  } } =\dfrac{1}{2} D^{-1} (x-y) \label{proJ}
\end{equation}
and 
\begin{equation}
<K_{-}(x) K_{-}(y)> \equiv \dfrac{\int D K_{-} K_{-}(x) K_{-}(y) 
e^{ -\int d^d x d^d y \tfrac{1}{4} K_{-} (x) W(x-y) K_{-}(y) }  }{\int D K_{-}  
e^{ -\int d^d x d^d y \tfrac{1}{4} K_{-} (x) W(x-y) K_{-}(y)  } } =2 A^{-1} (x-y) .  \label{proK}
\end{equation}
Then we can expand the $E_{\Omega}$ for $n=2$ as 
\begin{equation}
\begin{split}
&E_{\Omega}= \mathrm{Tr} \rho_{0\Omega}^2 
[1-2 \int d^d x d^d y  <J_{-}(x) J_{-}(y)> :\phi_{-}(x) \phi_{-}(y) :          \\
&-\dfrac{1}{2} \int d^d x d^d y  <K_{-}(x) K_{-}(y)> :\pi_{-}(x) \pi_{-}(y) : + \dots ]   .
\end{split}  \label{expansion normal opjkf n=2}
\end{equation}

Next let us apply above results to  
the mutual R\'{e}nyi information  
$I^{(2)}(A,B) $ of
disjoint compact spatial regions A and B in the vacuum states 
of the massless free scalar field.   
We express $E_{\Omega}$ as a sum of the local operators 
at some conventionally chosen points $(r_{A}, r_{B})$ inside $A$ and $B$.  
From (\ref{expansion normal opjkf n=2}), we have 
\begin{equation}
\begin{split}
&E_{A(B)}= \mathrm{Tr} \rho_{0A(B)}^2 
[1-2  : \phi_{-}^2 (r_{A(B)}) :  C_{A(B)} + \dots ]
\end{split}  \label{EA(B) massless}
\end{equation}
where 
\begin{equation}
\begin{split}
C_{A(B)}=  \int d^d x d^d y  <J_{-}(x) J_{-}(y)> .
\end{split}  \label{CA(B)}
\end{equation}
From the local property (\ref{property2}), $E_{A\cup B}=E_A E_B$, and  
(\ref{EA(B) massless}) we obtain   
\begin{equation}
\begin{split}
&\dfrac{\mathrm{Tr} \rho_{0A\cup B}^2}{\mathrm{Tr} \rho_{0A}^2 \mathrm{Tr} \rho_{0B}^2} 
=\dfrac{\bra{0^{(2)}} E_A E_B \ket{0^{(2)}}   }{ \bra{0^{(2)}} E_A \ket{0^{(2)}} \bra{0^{(2)}} E_B \ket{0^{(2)}}  }  
\simeq 1+\dfrac{1}{2} C_{A} C_{B} (W^{-1}(r) )^2   ,  
\end{split}  \label{mutual massless vacuum}
\end{equation}
where we have used 
$\bra{0^{(2)}} : \phi_{-}^2 (x) : : \phi_{-}^2 (y) : \ket{0^{(2)}} = 
2 (\bra{0}\phi(x) \phi(y) \ket{0})^2 =  \tfrac{1}{2} (W^{-1}(x-y) )^2  $. 
We use the notation $E_{p}=|p|$ and 
\begin{equation}
W^{-1}(x-y)= \int \dfrac{d^d p}{(2\pi)^d} \dfrac{1}{|p|}  e^{ip(x-y)} =\dfrac{B_{d}}{|x-y|^{d-1}} ,
\end{equation}
where $B_{d}$ is a numerical constant. 
Thus the the mutual R\'{e}nyi information  
$I^{(2)}(A,B) $ is 
\begin{equation}
\begin{split}
I^{(2)}(A,B) \simeq \dfrac{1}{2} C_{A} C_{B} \dfrac{B_d^2}{r^{2d-2}}  .  
\end{split}  \label{mutual massless vacuum2}
\end{equation}

We can compute $C_{A(B)}$ numerically at least. 
The expression (\ref{CA(B)}) of $C_{A(B)}$ is 
useful for numerical computation.  
Furthermore, we can obtain the alternative expression of $C_{A(B)}$ 
as follows. 
Let us consider the following generating function, 
\begin{equation}
\begin{split}
&G(\sigma) =\int DJ_{-} 
\exp [ -\int d^d x d^d y J_{-} (x) W^{-1}(x-y) J_{-}(y) 
+i\sigma \int d^d x J_{-}(x) ]   ,
\end{split}  \label{Generating}
\end{equation}
where $\sigma$ is a c number.    
We can obtain the coefficients of $:\phi^{2m}_{-}(r_A):$ 
in the expansion of $E_A$  
from $G(\sigma)$.   
For example, we have  
\begin{equation}
\begin{split}
C_{A}=\dfrac{-1}{G(0)} 
\left. \dfrac{\partial^2 G}{\partial \sigma^2} \right|_{\sigma=0} .
\end{split}  \label{Generating CA}
\end{equation}
We can rewrite $G(\sigma)$ by using the $(d+1)$ dim Euclidean path integral as 
\begin{equation}
\begin{split}
&G(\sigma) =\dfrac{1}{Z(0)} \int DJ_{-} 
Z(J_{-}) \exp [ i\sigma \int d^d x J_{-}(x) ]   ,
\end{split}  \label{Generating2}
\end{equation}
where 
\begin{equation}
\begin{split}
&Z(J_{-})=\int D\varphi 
\exp [-\dfrac{1}{2} \int d^{d+1} x (\partial_{\mu} \varphi)^2
-2i\int d^d x J_{-}\varphi|_{\tau =0}   ]   ,
\end{split}  \label{Z}
\end{equation}
where $\tau$ is the Euclidean time coordinate and 
$\varphi(\tau, x)$ is a scalar field in $(d+1)$ dimensional 
Euclidean space.  
First, we perform the $J_{-}$ integral and obtain 
\begin{equation}
\begin{split}
&G(\sigma) =\dfrac{1}{Z(0)} \int D \varphi 
\exp [-\dfrac{1}{2} \int d^{d+1} x (\partial_{\mu} \varphi)^2 ] 
\prod_{x\in \Omega} \delta (\sigma -2 \varphi (\tau=0, x)) .
\end{split}  \label{Generating3}
\end{equation}
We decompose $\varphi$ into the quantum part $\varphi_{q}$ 
and the classical part $\varphi_{cl}$. 
The $\varphi_{cl}$ is the solution of $\partial^2 \varphi =0$ 
and satisfy the boundary condition, 
\begin{equation}
\begin{split}
\varphi_{cl}(\tau=0, x)=\dfrac{\sigma}{2} ~~~(x \in A),~ \text{and}
~~~\varphi_{cl}(\tau, x)=0 ~~~(|\tau^2+x^2| \rightarrow \infty).
\end{split}  \label{Generating3}
\end{equation}
Thus the region $(\tau=0, x\in A)$ acts like a conductor 
where electrostatic potential is $\sigma/2$ and 
$\varphi_{cl}(\tau,x)$ is electrostatic potential at $(\tau,x)$.
We perform the $\varphi_q$ integral and obtain  
\begin{equation}
\begin{split}
&G(\sigma) =G(0) \exp [-\dfrac{1}{2} \int d^{d+1} x (\partial_{\mu} \varphi_{cl})^2 ] 
=G(0) \exp [-\dfrac{\sigma^2}{8} \mathbf{C}_{A}  ] , 
\end{split}  \label{Generating4}
\end{equation}
where $\dfrac{1}{2} \int d^{d+1} x (\partial_{\mu} \varphi_{cl})^2 $ is 
the electrostatic energy and we have rewritten it by using 
$\mathbf{C}_{A}$ which is electrostatic capacitance of 
the conductor $(\tau=0, x\in A)$. 
From (\ref{Generating CA}) and (\ref{Generating4}) we have 
\begin{equation}
\begin{split}
C_{A}=\dfrac{1}{4} \mathbf{C}_{A}   . 
\end{split}  \label{Generating5}
\end{equation}
Thus, from (\ref{mutual massless vacuum2}) and (\ref{Generating5}), we reproduced the result that $I^{(2)}(A,B)$ in the vacuum state is proportional to the product of 
the electrostatic capacitance of each regions obtained by Cardy \cite{Ca1}.

\subsection{The mutual information of locally excited states }
We consider the mutual R\'{e}nyi information  
$I^{(n)}(A,B) = S_{A}^{(n)} +S_{B}^{(n)} -S_{A\cup B}^{(n)}$ of
disjoint compact spatial regions A and B in the locally excited states of 
the (d+1) dimensional free massless scalar field theory 
in the limit when the separation $r$ between $A$ and
$B$ is much greater than their sizes $R_{A,B}$. 

We consider the following locally excited state,
\begin{equation}
\ket{\Psi} =N (O_{iA} O_{jB} +O_{i'A} O_{j'B} ) \ket{0}, 
\label{excited state}
\end{equation} 
where $N$ is a real normalization constant and
 $O_{iA}$ and $O_{i'A}$ ($O_{jB}$ and $O_{j'B}$) are operators on $A$ ($B$) 
and $i$ and $i'$ ($j$ and $j'$) label a kind of operators. 
We impose following orthogonal conditions, 
\begin{equation}
\bra{0} O_{iA}^{\dagger} O_{i'A} \ket{0}  =\bra{0} O_{jB}^{\dagger} O_{j'B} \ket{0}=0 . \label{orthogonal}
\end{equation}
This state is similar to the EPR state. 
We compute the mutual R\'{e}nyi information of this state.  
When $r\rightarrow \infty$, the mutual information $I^{(n)}(A,B)$ 
is the same as that of the general QFT which have a mass gap in 
(\ref{mutual Renyi mass gap}). 
In the mass gap case there is a correction which is $O(e^{-mR_{A(B)}})$
for (\ref{mutual Renyi mass gap}). 
In the free massless case there is a correction which is $O(1/R_{A(B)})$.  

We consider the leading term of $I^{(n)}(A,B)$ which depends on $r$ 
for $r \gg R_{A},R_{B}$. 
Because the symmetric property (\ref{property1}) is important 
to determine the $r$ dependence of $I^{(n)}(A,B)$ as we will show it later,   
from now on, we impose the condition that 
under the sign changing  transformation 
$(\phi, \pi)\rightarrow (-\phi, -\pi)$  
the operators $O$ in (\ref{excited state}) is transformed as 
\begin{equation}
O \rightarrow (-1)^{|O|} O ,  \label{sign change}
\end{equation}
where $|O|=0~ \text{or}~ 1$.  
From the normalization condition $\braket{\Psi|\Psi}=1$ we have  
\begin{equation}
N^{-2} = \bra{0} [ O_{iA}^{\dagger} O_{iA}O_{jB}^{\dagger} O_{jB}
+  O_{iA}^{\dagger} O_{i'A}O_{jB}^{\dagger} O_{j'B} 
+ O_{i'A}^{\dagger} O_{iA}O_{j'B}^{\dagger} O_{jB} 
+  O_{i'A}^{\dagger} O_{i'A}O_{j'B}^{\dagger} O_{j'B}] \ket{0} . \label{normalization massless}
\end{equation}
By using (\ref{formula}), we obtain $\mathrm{Tr} \rho_{\Omega}^n$ for $\rho=\ket{\Psi}\bra{\Psi}$ as 
\begin{equation}
\begin{split}
&\mathrm{Tr} \rho_{\Omega}^n = \bra{\Psi^{(n)}} E_{\Omega} \ket{\Psi^{(n)}}  \\
&=N^{2n} \bra{0^{(n)}}  (O_{iA}^\dagger O_{jB}^\dagger +O_{i'A}^\dagger O_{j'B}^\dagger ) ^{(1)} \dots 
 (O_{iA}^\dagger O_{jB}^\dagger +O_{i'A}^\dagger O_{j'B}^\dagger ) ^{(n)}  \\
& \times E_{\Omega} 
  (O_{iA} O_{jB} +O_{i'A} O_{j'B} ) ^{(1)} \dots 
 (O_{iA} O_{jB} +O_{i'A} O_{j'B} ) ^{(n)} \ket{0^{(n)}} , 
\label{omega massless}
\end{split}
\end{equation}
where $O^{(l)}$ are operators in the Hilbert space of the $l$-th copy 
and the $\Omega$ is $A, B$ or $A\cup B$.

\subsubsection{The case $O_{i'A}=O_{j'B}=0$}
First we consider the case $O_{i'A}=O_{j'B}=0$ and $O_{iA}$ and $O_{jB}$ 
are nonzero
\begin{equation}
\ket{\Psi_1} =N_1 O_{iA} O_{jB} \ket{0}. 
\label{excited state1}
\end{equation}   
From the normalization condition we have 
\begin{equation}
N_{1}^{-2} = \bra{0}  O_{iA}^{\dagger} O_{iA}O_{jB}^{\dagger} O_{jB}
\ket{0} . \label{normalization massless1}
\end{equation}
From the condition (\ref{sign change}), 
$|O_{iA}^{\dagger} O_{iA}|=0$ and the OPE of $O_{iA}^{\dagger} O_{iA}$ 
has the form 
\begin{equation}
O_{iA}^{\dagger} O_{iA}
=\bra{0}O_{iA}^{\dagger} O_{iA} \ket{0}[1+ 
C_{iA}^{:\phi^2:} :\phi^2:(r_A)+\dots ] , \label{OPE ii}
\end{equation}
where 
$(r_{A}, r_{B})$ are 
some conventionally chosen points inside $A$ and $B$. 
The key point is that there is not $\phi$ in the OPE (\ref{OPE ii}) 
because of $|O_{iA}^{\dagger} O_{iA}|=0$. 
The OPE of $O_{jB}^{\dagger} O_{jB}$ has the same form as that of 
$O_{iA}^{\dagger} O_{iA}$ 
because of $|O_{jB}^{\dagger} O_{jB}|=0$. 
Thus we have 
\begin{equation}
N_{1}^{-2} = \bra{0}  O_{iA}^{\dagger} O_{iA} 
\ket{0} \bra{0} O_{jB}^{\dagger} O_{jB}
\ket{0}[1+O(1/r^{2d-2})] , \label{normalization massless1-2}
\end{equation}
where we have used 
$\bra{0}:\phi^2(r_A)::\phi^2(r_B): \ket{0}\propto 1/r^{2d-2}$. 

We consider $\mathrm{Tr} \rho_{A}^n$. 
From (\ref{omega massless}) we have 
\begin{equation}
\begin{split}
&\mathrm{Tr} \rho_{A}^n = \bra{\Psi_{1}^{(n)}} E_{\Omega} \ket{\Psi_{1}^{(n)}}  \\
&=N_{1}^{2n} \bra{0^{(n)}}  (O_{iA}^\dagger O_{jB}^\dagger ) ^{(1)} \dots 
 (O_{iA}^\dagger O_{jB}^\dagger ) ^{(n)}  E_{A} 
  (O_{iA} O_{jB}  ) ^{(1)} \dots 
 (O_{iA} O_{jB}  ) ^{(n)} \ket{0^{(n)}} \\ 
&=N_{1}^{2n} \bra{0^{(n)}}  O_{iA}^{(1)\dagger}  \dots 
O_{iA}^{(n)\dagger}  E_{A} 
  O_{iA}^{(1)} \dots 
 O_{iA}^{(n)} \cdot
O_{jB}^{(1)\dagger} \dots O_{jB}^{(n)\dagger} 
O_{jB}^{(1)} \dots O_{jB}^{(n)}  \ket{0^{(n)}}  . 
\label{A massless1}
\end{split}
\end{equation}
From (\ref{property1}) and (\ref{sign change}), 
$O_{iA}^{(1)\dagger}  \dots 
O_{iA}^{(n)\dagger}  E_{A} 
  O_{iA}^{(1)} \dots 
 O_{iA}^{(n)}$ and 
$O_{jB}^{(1)\dagger} \dots O_{jB}^{(n)\dagger} 
O_{jB}^{(1)} \dots O_{jB}^{(n)} $ are even 
under the sign changing transformation 
\begin{equation}
\begin{split}
(\phi^{(1)}, \dots , \phi^{(n)}, \pi^{(1)}, \dots , \pi^{(n)} )
\rightarrow 
(-\phi^{(1)}, \dots , -\phi^{(n)}, -\pi^{(1)}, \dots , -\pi^{(n)} ).
\end{split} \label{parity}
\end{equation}
So the OPEs of these operators are  
\begin{equation}
\begin{split}
O_{iA}^{(1)\dagger}  \dots 
O_{iA}^{(n)\dagger}  E_{A} 
  O_{iA}^{(1)} \dots 
 O_{iA}^{(n)}
=\bra{0} O_{iA}^{(1)\dagger}  \dots 
O_{iA}^{(n)\dagger}  E_{A} 
  O_{iA}^{(1)} \dots 
 O_{iA}^{(n)}\ket{0}
[1+\sum_{l,m}\tilde{C}_{iA}^{:\phi^{(l)}\phi^{(m)}:}:\phi^{(l)}\phi^{(m)}:(r_A)+\dots] 
\end{split} \label{OPE 2}
\end{equation}
and 
\begin{equation}
\begin{split}
O_{jB}^{(1)\dagger} \dots O_{jB}^{(n)\dagger} 
O_{jB}^{(1)} \dots O_{jB}^{(n)}
=\bra{0} O_{jB}^{(1)\dagger} \dots O_{jB}^{(n)\dagger} 
O_{jB}^{(1)} \dots O_{jB}^{(n)} \ket{0}
[1+\sum_{l}\tilde{D}_{jB}^{:\phi^{(l)}\phi^{(l)}:}:\phi^{(l)}\phi^{(l)}:(r_B)+\dots] .
\end{split} \label{OPE 3}
\end{equation}
Thus, we substitute (\ref{normalization massless1-2}), 
(\ref{OPE 2}) and (\ref{OPE 3}) into (\ref{A massless1}) and obtain 
the power of $r$ of the subleading term of $\mathrm{Tr} \rho_{A}^n$ 
\begin{equation}
\begin{split}
\mathrm{Tr} \rho_{A}^n = 
\mathrm{Tr} \rho_{A}^n |_{r \rightarrow \infty} [1+O(1/r^{2d-2})] . 
\label{A massless1-2}
\end{split}
\end{equation}
In the same way, 
we can obtain the power of $r$ of the subleading terms of $\mathrm{Tr} \rho_{B}^n$ and 
$\mathrm{Tr} \rho_{A\cup B}^n$ 
\begin{equation}
\begin{split}
\mathrm{Tr} \rho_{B}^n = 
\mathrm{Tr} \rho_{B}^n |_{r \rightarrow \infty} [1+O(1/r^{2d-2})] , ~~~
\mathrm{Tr} \rho_{A\cup B}^n = 
\mathrm{Tr} \rho_{A\cup B}^n |_{r \rightarrow \infty} [1+O(1/r^{2d-2})] .
\label{B and AB massless1-2}
\end{split}
\end{equation}
Thus the mutual R\'{e}nyi information is 
\begin{equation}
\begin{split}
I^{(n)}(A,B)=
O(1/r^{2d-2}) ,
\label{massless mutual 1}
\end{split}
\end{equation}
where we have used $I^{(n)}(A,B)|_{r \rightarrow \infty}=0$ for 
$O_{i'A}=O_{j'B}=0$ from (\ref{mutual Renyi mass gap}).

\subsubsection{The case $O_{iA}, O_{jB}, O_{i'A}~ \text{and}~ O_{j'B}$ are nonzero}
There are two cases which are different powers of $r$ 
of the subleading term of $I^{(n)}(A,B)$.

(i)The case $|O_{iA}|=|O_{i'A}|$

In this case the operators at $A$ in the expansions of 
$N^{-2},~\mathrm{Tr} \rho_{\Omega}^n ~(\Omega =A,B,A\cup B) $ 
in (\ref{normalization massless}) and (\ref{omega massless})
are even under the sign changing transformation ($\ref{parity}$). 
So the subleading terms of them come from the operators $:\phi^{(l)} \phi^{(m)}: (r_{A})$. 
Thus the mutual R\'{e}nyi information is 
\begin{equation}
\begin{split}
I^{(n)}(A,B)=I^{(n)}(A,B)|_{r\rightarrow \infty}+
O(1/r^{2d-2}) ,
\label{massless mutual 2}
\end{split}
\end{equation}
where $I^{(n)}(A,B)|_{r\rightarrow \infty}$ is the same form as  (\ref{mutual Renyi mass gap}).

(ii)The case $|O_{iA}| \neq |O_{i'A}|$ and $|O_{jB}| \neq |O_{j'B}|$

There are some terms which are odd under the sign changing transformation ($\ref{parity}$).  
As an example, in the expansion of $\mathrm{Tr} \rho_{A\cup B}^n$  
in (\ref{omega massless}), 
we consider the following term, 
\begin{equation}
\bra{0^{(n)}} O_{i'A}^{(1)\dagger} O_{iA}^{(2)\dagger} \dots O_{iA}^{(n)\dagger} E_A 
O_{iA}^{(1)} O_{iA}^{(2)} \dots O_{iA}^{(n)} 
\cdot O_{j'B}^{(1)\dagger} O_{jB}^{(2)\dagger} \dots O_{jB}^{(n)\dagger} E_B 
O_{jB}^{(1)} O_{jB}^{(2)} \dots O_{jB}^{(n)} \ket{0^{(n)}}.  
\label{example massless}
\end{equation}
In (\ref{example massless}), the operators at $A$ and at $B$ 
are both odd under the sign changing transformation ($\ref{parity}$). 
So the leading $r$ dependent term of it come from the operators $\phi^{(l)}$. 
Thus the mutual R\'{e}nyi information is 
\begin{equation}
\begin{split}
I^{(n)}(A,B)=I^{(n)}(A,B)|_{r\rightarrow \infty}+
O(1/r^{d-1}) ,
\label{massless mutual 3}
\end{split}
\end{equation}
where $I^{(n)}(A,B)|_{r\rightarrow \infty}$ is the same form as  (\ref{mutual Renyi mass gap}).

\section{Conclusion and discussions}

We developed the computational method of R\'{e}nyi 
entanglement entropy based on the idea that 
$\mathrm{Tr} \rho_{\Omega}^n$ 
is written  
as the expectation value of the local operator at $\Omega$. 
We expressed $\mathrm{Tr} \rho_{\Omega}^n$ as 
the expectation value of the glueing operator $E_{\Omega}$,  
$\mathrm{Tr} \rho_{\Omega}^n =\mathrm{Tr}(\rho^{(n)}E_{\Omega})$. 
We constructed explicitly $E_{\Omega}$ and 
 investigated its general properties. 
For a free scalar field, we rewrote $E_{\Omega}$ in (\ref{opjkf}) 
using the normal ordering. 
In the case $n=2$, we obtained a simple expression of  $E_{\Omega}$ 
and reproduced the result that $I^{(2)}(A,B)$ in the vacuum state is proportional to the product of 
the electrostatic capacitance of each regions obtained by Cardy \cite{Ca1}. 
The coefficients of the expansion of $E_{\Omega}$ is obtained by the propagators of 
$J_{-}$ and $K_{-}$ in (\ref{proJ}) and (\ref{proK}). 
We can compute these propagators numerically at least and the expression (\ref{normal opjkf n=2}) 
is useful for numerical calculation. 

The advantages of this methods are that we can use ordinary technique in QFT such as 
OPE and the cluster decomposition property and 
that we can use the general properties and the explicit expression of the glueing operator to 
compute systematically the  R\'{e}nyi entropy for an arbitrary state.


We applied this method to consider the mutual R\'{e}nyi information 
$I^{(n)}(A,B)$ 
of disjoint compact spatial regions $A$ and $B$ in the locally excited states defined by acting 
the local operators at $A$ and $B$ on the vacuum 
of a $(d+1)$-dimensional field theory, 
in the limit when the separation $r$ between $A$ and $B$ is much greater than their sizes $R_{A,B}$. 
For 
the general QFT which has a mass gap, 
we computed $I^{(n)}(A,B)$ explicitly and find that this result is interpreted in terms of an entangled state 
in quantum mechanics. 
Interestingly, the mutual information in the QFT measures only the quantum entanglement  
in the limit $r \rightarrow \infty$ although 
the mutual information measures generally 
the total of the quantum entanglement and the classical one \cite{Ni}. 
So, in this limit, the mutual information is a good measure of quantum entanglement in this sense. 
For a free massless scalar field, 
we showed that for some classes of excited states 
$I^{(n)}(A,B)-I^{(n)}(A,B)|_{r \rightarrow \infty} =C^{(n)}_{AB}/r^{\alpha (d-1)}$ 
where $\alpha=1~ \text{or}~ 2$ which is determined by the property of the local operators 
under the transformation $\phi \rightarrow  -\phi$ and $\alpha=2$ for the vacuum state. 

Finally we discuss the generalization of our method. 
Although we considered only the locally excited states, 
we can apply our method to more general excited states, 
for example, many particle states and thermal states.  
We might be able to generalize our method to fermionic fields. 
We could apply our method to perturbative calculation in an interacting field theory.


\acknowledgments

I am grateful to Tadashi Takayanagi for a careful reading of this manuscript 
and useful comments and discussions. 
I also would like to thank Pawel Caputa, John Cardy, Song He, Masahiro Nozaki, 
Tokiro Numasawa, and Kento Watanabe for useful discussions. 
This work is supported by JSPS Grant-in-Aid for Scientific 
Research (B) No.25287058 and JSPS Grant-in-Aid for Challenging 
Exploratory Research No.24654057.

\appendix

\section{R\'{e}nyi entropy for the vacuum state in free scalar field theory}
In this appendix we show that 
 (\ref{vacuum}) reproduces the same result as that of 
the real time approach \cite{Bombelli:1986rw, Sr, CW}
 which is based on the wave functional calculation.  

From (\ref{tilde S}) and (\ref{vacuum}), 
we perform the $K$ integral in  (\ref{vacuum}) and obtain 
\begin{equation}
\mathrm{Tr} \rho_{0\Omega}^n 
= \left( Det \left( \dfrac{A}{4\pi} \right) \right) ^{-n/2}
\int \prod_{j=1}^{n} \prod_{x \in \Omega} DJ^{(j)}(x) \exp [ 
-\int d^{d} x d^{d} y  (J^{(1)}(x), \cdots, J^{(n)}(x)) M_n (x,y)
\begin{pmatrix}
             J^{(1)}(y)  \\
            \vdots \\
             J^{(n)}(y)
             \end{pmatrix} ] , \label{vacuum2}
\end{equation}
where 
\begin{equation}
M_n =
\begin{pmatrix}
            X & Y & 0 & \cdots &0 &Y   \\
            Y& X&Y&\cdots &0&0 \\
             0&Y& X&\cdots &0&0 \\
            \vdots &\vdots &\vdots & &  \vdots &\vdots \\
            0&0&0&\cdots &X&Y \\
             Y&0&0&\cdots &Y&X \\
             \end{pmatrix} , \label{M_n}
\end{equation}
here 
\begin{equation}
X=\dfrac{1}{2}(A^{-1}+D), ~~~ Y=\dfrac{1}{4}(A^{-1}-D)  \label{xy}
\end{equation}
and we have used the matrix notation in (\ref{matw}) and (\ref{matwin}).

The following calculation is analogous to that in \cite{CW}. 
From (\ref{vacuum2}) and (\ref{M_n}) we obtain 
\begin{equation}
\mathrm{Tr} \rho_{0\Omega}^n 
=\left( Det \left( \dfrac{A}{4\pi} \right) \right)^{-n/2} 
(Det (4\pi M_n) )^{-1/2},
 \label{vacuum3}
\end{equation}
where we have used the normalization condition of the $J$ integral 

$\int DJ^{(j) } \exp [-\int d^d x d^d y J^{(j)} (x) M(x,y) J^{(j)} (y) ] =(Det(4\pi M))^{-1/2} $ .

We rewrite $M_n$ as
\begin{equation}
 M_n = \dfrac{X}{2} \tilde{M_n}, \label{M_n2}
\end{equation}
where
\begin{equation}
\tilde{M_n} =
\begin{pmatrix}
            2 & -Z & 0 & \cdots &0 &-Z   \\
            -Z&2&-Z&\cdots &0&0 \\
             0&-Z&2&\cdots &0&0 \\
            \vdots &\vdots &\vdots & &  \vdots &\vdots \\
            0&0&0&\cdots &2&-Z \\
             -Z&0&0&\cdots &-Z&2 \\
             \end{pmatrix} , \label{tilde M_n}
\end{equation}
here
\begin{equation}
Z=-2X^{-1} Y =(1+AD)^{-1} (AD-1)  . \label{matz}
\end{equation}
We diagonalize $Z$ and denote the eigenvalues of $Z$ as $z_i$.
And we can diagonalize $\tilde{M_n}$ by Fourier transformation and obtain
\begin{equation}
Det \tilde{M_n} =
\prod_i \prod_{r=1}^n [2-2z_i \cos (\dfrac{2 \pi r}{n}) ] =\prod_i 2^n \dfrac{(1-\xi_i^n)^2}{(1+\xi_i^2)^n},  \label{det tilde M_n}
\end{equation}
where $\xi_i$ is defined as
\begin{equation}
z_i=2\xi_i/(\xi_i^2+1). \label{xi}
\end{equation}

From (\ref{vacuum3}) and (\ref{det tilde M_n}) we obtain
\begin{equation}
\mathrm{Tr} \rho_{0\Omega}^n  = \prod_{i} \dfrac{(1-\xi_i)^n }{(1-\xi_i^n)} .
\label{vacuum4}
\end{equation}
Thus we obtain the R\'{e}nyi entropies $S_{0\Omega}^{(n)}=(1-n)^{-1} \ln \mathrm{Tr} \rho_{0\Omega}^n$
and the entanglement entropy
$S_{0\Omega}=-\mathrm{Tr} \rho_{0\Omega} \ln \rho_{0\Omega} =
-\tfrac{\partial}{\partial n} \ln \mathrm{Tr} \rho_{0\Omega}^n |_{n=1}$
as follows:
\begin{align}
&S_{0\Omega}^{(n)}   =
\sum_i (1-n)^{-1} [n \ln (1-\xi_i)-\ln(1-\xi_i^n) ] ,     \label{vacuum r entropy} \\
&S_{0\Omega}   =
\sum_i [- \ln (1-\xi_i)-\dfrac{\xi_i}{1-\xi_i} \ln \xi_i ]
 .\label{vacuum entropy}
\end{align}

In order to show that 
 (\ref{vacuum r entropy}) and (\ref{vacuum  entropy})  are the same results as those of 
the real time approach, 
we rewrite  (\ref{vacuum r entropy}) and(\ref{vacuum  entropy}). 
We define the matrix  
\begin{equation}
\tilde{C}=\dfrac{1}{2} (DA)^{1/2} ,
\end{equation}
and rewrite the matrix $Z$ in (\ref{matz}) as 
\begin{equation}
Z=(1+4(\tilde{C}^{T})^2)^{-1} (4(\tilde{C}^{T})^2 -1) . \label{matz2}
\end{equation}
From (\ref{xi}) and (\ref{matz2}) we obtain 
\begin{equation}
\tilde{C}^{T}_i =\dfrac{1}{2} (1+\xi_i) (1-\xi_i)^{-1}.  \label{tilde C}
\end{equation}
From (\ref{tilde C}), we rewrite   (\ref{vacuum r entropy}) and(\ref{vacuum  entropy}) as 
\begin{equation}
\begin{split}
S_{0\Omega}^{(n)}   &= 
\sum_i (n-1)^{-1} \ln [ (\tilde{C}^{T}_i+1/2)^{n} -  (\tilde{C}^{T}_i-1/2)^{n}]  \\
&=(n-1)^{-1} \mathrm{tr} \ln [ (\tilde{C}^{T}+1/2)^{n} -  (\tilde{C}^{T}-1/2)^{n}] \\
&=(n-1)^{-1} \mathrm{tr} \ln [ (\tilde{C}+1/2)^{n} -  (\tilde{C}-1/2)^{n}] ,
\end{split}   \label{vacuum r entropy2}
\end{equation}
\begin{equation}
\begin{split}
S_{0\Omega}   
=\mathrm{tr} [ (\tilde{C}+1/2) \ln  (\tilde{C}+1/2)  -  (\tilde{C}-1/2) \ln (\tilde{C}-1/2)] .
\end{split}  \label{vacuum entropy2}
\end{equation}
 (\ref{vacuum r entropy2}) and (\ref{vacuum  entropy2})  are the same results as those of 
the real time approach (see e.g. \cite{CaHu2}).






\begin{thebibliography}{99}

\bibitem{Bombelli:1986rw}
  L.~Bombelli, R.~K.~Koul, J.~Lee and R.~D.~Sorkin,
  ``A Quantum Source of Entropy for Black Holes,''
  Phys.\ Rev.\ D {\bf 34} (1986) 373.

\bibitem{Sr}
   M.~Srednicki,
   ``Entropy and area,''
   Phys.\ Rev.\ Lett.\  {\bf 71} (1993) 666  [hep-th/9303048].



\bibitem{Ereview}
  J.~Eisert, M.~Cramer and M.~B.~Plenio,
  ``Area laws for the entanglement entropy - a review,''
  Rev.\ Mod.\ Phys.\  {\bf 82} (2010) 277  [arXiv:0808.3773 [quant-ph]].


\bibitem{La}
  J.I.Latorre and A.Riera, 
  ``A short review on entanglement in quantum spin systems,'' 
  J. Phys. A 42 (2009) 4002 [arXiv:0906.1499 [hep-th]] 




\bibitem{ShTa}
  N.~Shiba and T.~Takayanagi,
  ``Volume Law for the Entanglement Entropy in Non-local QFTs,''  arXiv:1311.1643 [hep-th].  



\bibitem{Ka}
  J.~L.~Karczmarek and P.~Sabella-Garnier,
  ``Entanglement entropy on the fuzzy sphere,''  arXiv:1310.8345 [hep-th].  



\bibitem{FS1}
M. M. Wolf, ``Violation of the entropic area law for fermions'', Phys. Rev. Lett. 96 (2006)
010404 [quant-ph/0503219]; 
D. Gioev, I. Klich, ``Entanglement entropy of fermions in any dimension and the Widom
conjecture,'' Phys. Rev. Lett. 96 (2006) 100503 [quant-ph/0504151].




\bibitem{Ca1}
J. Cardy, ``Some results on the mutual information of disjoint regions in higher dimensions,'' J.
Phys. A 46 (2013) 285402 [arXiv:1304.7985]



\bibitem{Ca2}

P. Calabrese, J. Cardy and E. Tonni, ``Entanglement entropy of two disjoint intervals in
conformal field theory II,'' J. Stat. Mech. 1101 (2011) P01021 [arXiv:1011.5482]



\bibitem{He}
M. Headrick, ``Entanglement Renyi entropies in holographic theories,'' Phys. Rev. D 82 (2010)
126010 [arXiv:1006.0047]




\bibitem{CaHu1}
H. Casini and M. Huerta, ``Remarks on the entanglement entropy for disconnected regions,''
JHEP 03 (2009) 048 [arXiv:0812.1773]

\bibitem{CaHu2}
H. Casini and M. Huerta, ``Entanglement entropy in free quantum field theory,''
J.Phys. A42 (2009) 504007, arXiv:0905.2562 [hep-th].



\bibitem{Sh1}
N. Shiba, ``Entanglement Entropy of Two Black Holes and Entanglement Entropic
Force,'' Phys.Rev. D83 (2011) 065002, arXiv:1011.3760 [hep-th].

\bibitem{Sh2}
N. Shiba, ``Entanglement Entropy of Two Spheres,'' JHEP 1207 (2012) 100,
arXiv:1201.4865 [hep-th].



\bibitem{Sch}
H.J.Schnitzer, "Mutual Renyi information for two disjoint compound systems", 
arXiv:1406.1161[hep-th].


\bibitem{Fa}
T. Faulkner, A. Lewkowycz and J. Maldacena, ``Quantum corrections to holographic
entanglement entropy," JHEP 1311, 074 (2013) [arXiv:1307.2892].






\bibitem{RT}
  S.~Ryu and T.~Takayanagi,
  ``Holographic derivation of entanglement entropy from AdS/CFT,''
  Phys.\ Rev.\ Lett.\  {\bf 96} (2006) 181602;
 ``Aspects of holographic entanglement entropy,''
  JHEP {\bf 0608} (2006) 045.
  




\bibitem{NNT}
M. Nozaki, T. Numasawa and T. Takayanagi, "Quantum Entanglement of Local Operators in Conformal Field Theories," Phys. Rev. Lett. 112, 111602 (2014)
[arXiv:1401.0539 [hep-th]].



\bibitem{No}
M. Nozaki, "Notes on Quantum Entanglement of Local Operators," arXiv:1405.5875
[hep-th].


\bibitem{HNTW}
S. He, T. Numasawa, T. Takayanagi and K. Watanabe, "Quantum Dimension as Entanglement
Entropy in 2D CFTs,"  arXiv:1403.0702 [hep-th].

\bibitem{CNT}
P. Caputa, M. Nozaki and T. Takayanagi, 
"Entanglement of Local Operators in large N CFTs," 
arXiv:1405.5946 [hep-th]



  
  
\bibitem{Ni}
M. Nielsen and I. Chuang, Quantum Computation and Quantum Information, Cambridge
University Press, Cambridge, England (2000), pg. 9.




\bibitem{CW}
C. G. Callan, Jr. and F. Wilczek, "On geometric entropy," Phys. Lett. B333,55 (1994),
arXiv:hep-th/9401072 .




\bibitem{Her}
C.P.Herzog, "Universal Thermal Corrections to Entanglement Entropy for Conformal Field Theories on Spheres," arXiv:1407.1358 [hep-th] .






















\end{thebibliography}
\end{document}